\documentclass[aps,prl,twocolumn,showpacs,superscriptaddress,groupedaddress]{revtex4-2}
\pdfoutput=1
\usepackage[caption=false]{subfig}
\usepackage{graphicx} 
\usepackage{mathrsfs}
\usepackage{amsfonts}
\usepackage{amsmath}
\usepackage[noabbrev]{cleveref}
\usepackage[toc,page]{appendix}
\usepackage{import}
\usepackage[nice]{units}
\usepackage{float}
\usepackage{color}

\begin{document}
\title{Recovering long-range cumulative response to geometric frustration in quasi-one-dimensional systems, mediated by constitutive softness}
\date{\today}
\author{Snir Meiri}
\email{snir.meiri@weizmann.ac.il} 
\author{Efi Efrati} 
\email{efi.efrati@weizmann.ac.il} 
\affiliation{Department of
Physics of Complex Systems, Weizmann Institute of Science, Rehovot
76100, Israel}

\begin{abstract} 
Cumulative geometric frustration can drive self-limited assembly and morphology selection through size-dependent energetic costs. However, the slenderness of quasi-one-dimensional systems generally suppresses the formation of long-range longitudinal gradients. We show that the suppression of longitudinal gradients can be overcome by tuning the ratio between the longitudinal and transverse (shear) moduli. We demonstrate the recovery of cumulative frustration across distinct quasi-one-dimensional systems, each frustrated through a different mechanism, by the introduction of a soft response mode. The resulting saturation length scale is universal and is dictated by the local constitutive information of the system. 
\end{abstract}
\maketitle

\begin{figure*}[htbp]
\includegraphics[width=16.4cm]{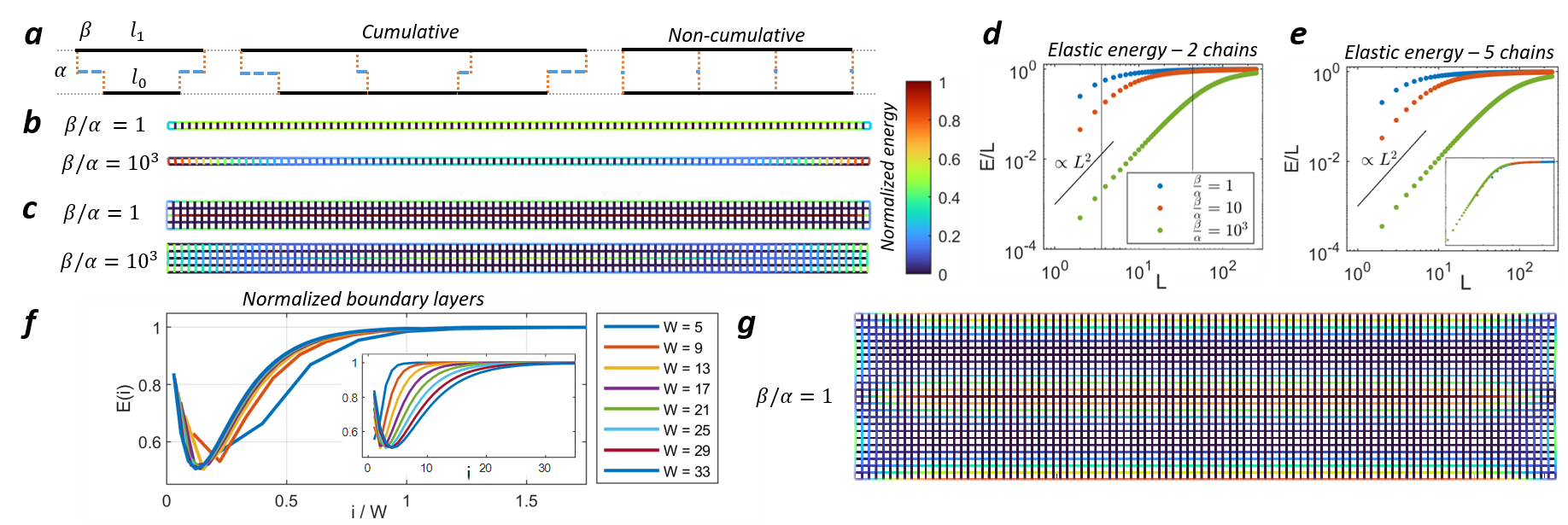}
\caption{Numerical results of coupled incommensurate chains of springs in $1D$. (a) Unit-cell sketch and schematic cumulative and non-cumulative solutions. Longitudinal springs of spring constant $\beta$ and rest-lengths $l_0$ and $l_1$ are shown in black lines. Coupling springs of spring constant $\alpha$ and vanishing rest-length are shown in blue dashed lines. The vertical spacing and orange dotted lines are for clarity. (b)-(c) Numerically minimized conformations for $L=100$ coupling sites at $\beta/\alpha=1$ and $10^3$. (b) Two chains with rest-lengths $1$ and $1.005$. (c) Five coupled spring chains with rest-lengths of $1.005,1.0025,1,1.0025$ and $1.005$. The color of each spring is indicative of its elastic energy, normalized to the spring of highest energy in the asymptotic solution. Vertical spacing between chains is for clarity and transverse edges represent coupling springs.  (d)-(e) Energy per unit length versus chain length for two and five chains with $\beta/\alpha=1,10,$ and $10^3$ (blue, orange, and green dots), normalized to asymptotic value of $1$. The left and right vertical lines in (d) mark the analytically calculated $\Tilde{L}$ for $\beta/\alpha=10$ and $10^3$. The inset in (e) shows the energy densities with the $x$-axis of each curve rescaled by $\sqrt{\beta/\alpha}$. (f)-(g) Scaling of the $\Tilde{L}$ with the number of chains, $W$. (f) Normalized longitudinal energy density near the left boundary versus the vertex index $i$ divided by $W$, color-coded according to $W$. The conformations have a symmetric linear rest-length profile about the midline, with $\Delta l=0.0025$ between chains, $\beta/\alpha=1$ and $L=100$.  The column energy $E(i)$ includes half the energy of the in-chain springs attached to column $i$, plus the tethering spring energy. The inset shows the energy density without normalization by $W$. (g) Representative minimal conformation for $W=25$, in the same format as (b) and (c).}  
\label{fig:incom}
\end{figure*}

Self-assembly is a key process in both naturally formed substances and synthetic systems. In some cases, the assembly process may vary depending on the size and shape of the already-formed structure. Such size-sensing significantly increases the span of attainable self-assembled structures. One rather natural manifestation of size-sensing in self-assembly processes is growth arrest, which is of high interest in various contexts, ranging from colloidal structures \cite{WWB+12,BET16b}, to DNA-origami \cite{Rot06} and protein-based nanomaterials \cite{KBS+14}. It can be obtained by forming closed structures \cite{HG21a,VHG+24}, such as a virus capsid, or by using many building blocks with specific interactions \cite{LKY05,ZMB14,KMM+25}. While the first approach significantly limits the types of attainable structures, the second requires high complexity for increasingly larger structures. 
Geometric frustration introduces unavoidable strains when assembling incompatible units and can thus act as a shape and size selective mechanism.
In general, the term geometric frustration describes systems in which the preferred local structure includes mutually contradicting tendencies that prevent its na\"ive realization as a bulk. Thus, when building blocks are assembled, the resulting structure must exhibit a compromise of local preferences. 
In cases where this compromise is spatially varying, leading to super-extensive energy scaling, the frustration is termed cumulative. This size and shape dependency of the energy scaling can lead to size-dependent phenomena such as growth arrest along different dimensions \cite{SG05,MPNM14,SHG22}, twist regulation \cite{AEKS11,HBBG16,HAS+19,ZGDS19,Gra20a,SDS+20} and emergence of ordered defect arrays \cite{Gra12,HAG23}. 
The spatially varying stress may also serve to modulate the growth of the assembly and lead to non-trivial morphological features. 
In contrast, non-cumulative frustration occurs when the optimal compromise adopted by misfitting units is spatially uniform, leading to extensive energy scaling. In such systems, geometric frustration cannot result in shape-dependent phenomena. We have recently proposed a framework that predicts the energy scaling exponent using the local constitutive information of the model \cite{ME21a}. Note that as super-extensivity cannot persist indefinitely, cumulative frustration can only be exhibited until a finite size $\Tilde{L}$, termed the frustration saturation scale, is met. 
Beyond this scale, system-specific frustration saturation mechanisms take over \cite{ME21a,HG21a,ME22b}, and the frustration becomes effectively non-cumulative. While the notion of a geometric charge typically requires the underlying system to be at least $2$-dimensional ($2D$), cumulative geometric frustration could also be realized in $1D$.
Recent works have studied self-limited quasi-$1D$ systems both experimentally and in simulations \cite{BWB+20b,THM+22,SHG24,WG24}. In this work, we study this saturation length-scale $\Tilde{L}$ in quasi-$1D$ frustrated systems and its dependence on their limited response mechanisms. Only systems in which $\Tilde{L}$ is comparable to or larger than their length may display longitudinal cumulative frustration.
Understanding what controls $\Tilde{L}$ may help identify natural systems in which geometric frustration plays a key role and help push $\Tilde{L}$ to larger scales in artificial systems. We proceed by examining four distinct quasi-$1D$ systems, frustrated by different mechanisms, yet demonstrating universal behavior for $\Tilde{L}$, and recovering cumulative frustration via constitutive softness.

To illustrate simply the interplay between geometric frustration and different response modes, we first consider the minimal system of two incommensurate spring chains restricted to a line. 
The Hamiltonian reads:
\begin{equation*}
H=\mathop\sum_{i=1}^{N-1}[\tfrac{\gamma }{2}(\Delta X_i-l_1)^2+\tfrac{\beta}{2}(\Delta x_i-l_0)^2]
+\mathop\sum_{i=1}^{N}\tfrac{\alpha}{2} (X_i-x_i)^2,    
\end{equation*}
where $\Delta X_i=X_{i+1}-X_i$, $\Delta x_i=x_{i+1}-x_i$, and $X_i$ ($x_i$) are the locations of the vertices in chains of springs of elastic constants $\gamma$ ($\beta$) and rest-lengths $l_1$ ($l_0$). Each pair ($X_i$,$x_i$) is coupled by a tethering spring of elastic constant $\alpha$ and vanishing rest-length. Having $l_0\neq l_1$ results in a conflict due to the tethering springs.
One can transform to local variables by defining $\Bar{l}=\frac{\gamma l_1+\beta l_0}{\gamma+\beta}$, $\eta_i=x_i-X_i$, $\xi_{i}=\frac{X_{i+1}-X_i}{\Bar{l}}$ and $\delta_i=\frac{x_{i+1}-x_i}{\Bar{l}}$. These variables are not independent, and must satisfy $\delta_i-\xi_i=\frac{\eta_{i+1}-\eta_i}{\Bar{l}}$. The Hamiltonian in the continuum limit reads:
\begin{align*}
H=\int_{-\frac{L}{2}}^{\frac{L}{2}}[\frac{\gamma \Bar{l}}{2}(\xi(x)-\frac{l_1}{\Bar{l}})^2
+\frac{\beta \Bar{l}}{2}(\delta(x)-\frac{l_0}{\Bar{l}})^2+\frac{\alpha}{2 \Bar{l}}\eta(x)^2]dx,
\end{align*}
alongside the compatibility condition $\delta-\xi=\eta'$, where the prime denotes the derivative with respect to the continuum longitudinal coordinate. From this condition, utilizing the approach presented in \cite{ME21a} we may predict the energy to scale super-extensively as $L^\lambda$, with $\lambda=3$. Indeed, allowing the fields to adopt uniform gradients results in a cumulative response whose energy scales as $\frac{\alpha  (l_0-l_1)^2}{24 \Bar{l}^3}L^3$. Comparing this energy to the optimal non-cumulative solution with vanishing gradients, $\frac{ (l_0-l_1)^2}{2\Bar{l} (\frac{1}{\beta} +\frac{1}{\gamma} )}L$, predicts their crossing at  $\Tilde{L} \sim \frac{\Bar{l}}{\sqrt{\alpha(\frac{1}{\beta}+\frac{1}{\gamma})}}$, defining the frustration saturation length-scale. This Hamiltonian can be solved exactly \cite{WG24,Le25}, and the obtained solution is in agreement with these results. The full solution for both the discrete and the continuous cases, as well as the analysis based on the compatibility condition, can be found in the supplementary material \cite{SupplementalMaterial}~\nocite{MATLAB,SN88}. Notice that due to the linearity of the compatibility condition, the rest-length mismatch, $(l_1-l_0)$, associated with the amplitude of the frustration, does not affect $\Tilde{L}$. 
Thus, in order to have a cumulative response over $\Tilde{L} \gg \Bar{l}$, the coupling strength between the chains must be much weaker than within the chains, i.e. $\alpha \ll \beta,\gamma$. 
In any other case, the cumulative response decays exponentially near the boundaries, with an associated decay length of the order of $\Bar{l}$. Even though this system is strictly $1D$, $\alpha$ couples the longitudinal response within the chains transversely, effectively playing the role of shear. Notice that in the limit of one chain being infinitely stiff, i.e. $\gamma\gg \beta,\alpha$, this system recovers the long-studied model of incommensurability \cite{FV49} relevant for understanding interfaces between crystals \cite{FV49,Bak82,RCIT14}, chains of DNA polybricks \cite{BWB+20b,WG24} and quantum-dot production \cite{BS12,JPW+19}. For the remainder of this manuscript we consider the symmetric case and set $\gamma=\beta$. This system can be generalized to $N$ sequentially coupled chains. Fig. \ref{fig:incom} shows the numerically obtained minimal-energy conformations for systems of two and five coupled chains for different values of $\beta/\alpha$. In both cases $\Tilde{L}$ scales as $\sqrt{\beta/\alpha}$, effectively reflecting the ratio of stretching to shear moduli. We also examined how $\Tilde{L}$ varies with the number of coupled chains, corresponding to the system’s effective width. Although the model is strictly $1D$ and lacks an intrinsic transverse length-scale,  the decay length of the boundary layer increases approximately linearly with this effective width. 

\begin{figure*}[htbp]
\includegraphics[width=16.4cm]{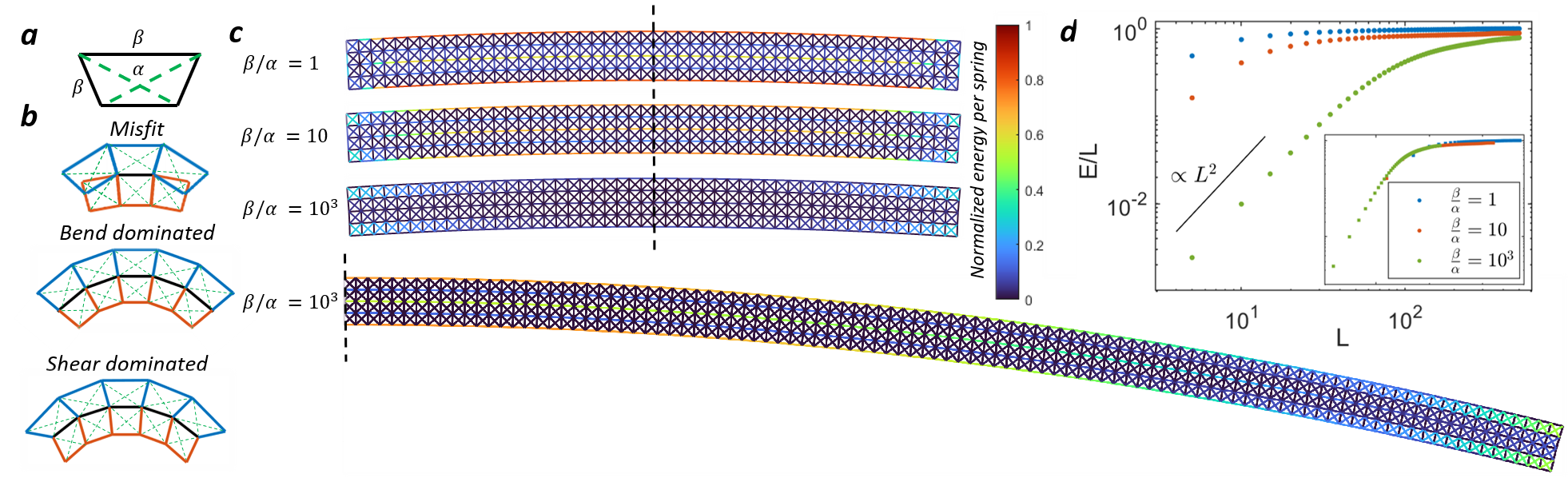}
\caption{Non-Euclidean spring chains restricted to the plane. (a) A single trapezoid, formed between two chains. Longitudinal and transverse springs of spring constant $\beta$ are shown in black lines, and diagonal springs of spring constant $\alpha$ are shown in green dashed lines. (b) Schematic representation of the source of frustration (top), and the invariant bend-dominated (middle) and cumulative shear-dominated (bottom) response modes. 
(c)-(d) Numerical results of five coupled chains with rest-length profile per chain (top to bottom) of $1.01$, $1.0056$, $1.0025$, $1.0006$ and $1$. Transverse springs are of rest-length $1$ and the diagonal springs have rest-lengths matching the regular trapezoid formed by the bounding springs at rest. (c) Minimal conformations for $L=50$ at $\beta/\alpha=1,10$ and $10^3$ and the right half of the symmetric minimal conformation for $L=200$ and $\beta/\alpha=10^3$. The color of each spring is indicative of its elastic energy according to the color-bar, normalized to the spring of highest energy in all the simulations. (d) Energy per unit length versus the length of the chain for $\beta/\alpha=1,10$ and $10^3$, marked as blue, orange and green dots, respectively. The energies are normalized to the maximal energy density. The inset shows the energy densities with the $x$-axis of each curve rescaled by $\sqrt{\beta/\alpha}$.
}
\label{fig:bend}
\end{figure*}

So far, the systems' response was restricted to $1D$. However, when embedded in higher-dimensional space the frustration may be partially or even fully alleviated. 
If one lifts the restriction to a line for the two coupled chains and separates the chains in the transverse direction, the frustration is lifted through the introduction of bending.
For an equi-spaced collection of curves in the plane it is straightforward to show that the radius of curvature must vary linearly with the transverse coordinate \cite{NE18}. 
Having a spatially invariant bent structure at constant transverse spacing requires a gradient of longitudinal length elements. Thus, coupling more than two chains transversely at equal spacing can lead to a stress-free conformation only for a linear rest-length profile. Any other profile, such as a quadratic one, will reintroduce frustration into the system. We thus simulate a $2D$ collection of $N$ chains, where the rest-lengths of the springs in the nth chain is set to $1+a n^2$, where $a$ is the amplitude of the frustration. We set  the elastic constant of both the longitudinal and the transverse springs to read $\beta$, and $\alpha$ is the elastic constant of the diagonal springs that control the shear modulus.
Fig. \ref{fig:bend} summarizes the numerical results of the discrete realization of such a quadratic profile. For $\beta\sim\alpha$, while the system is unmistakably frustrated, it only demonstrates longitudinal cumulative frustration over a narrow domain in the vicinity of the boundaries. As in the previous case, as the value of shear stiffness, $\alpha$, is diminished, long-lasting cumulative behavior is recovered. 

\begin{figure*}[htbp]
\includegraphics[width=16.4cm]{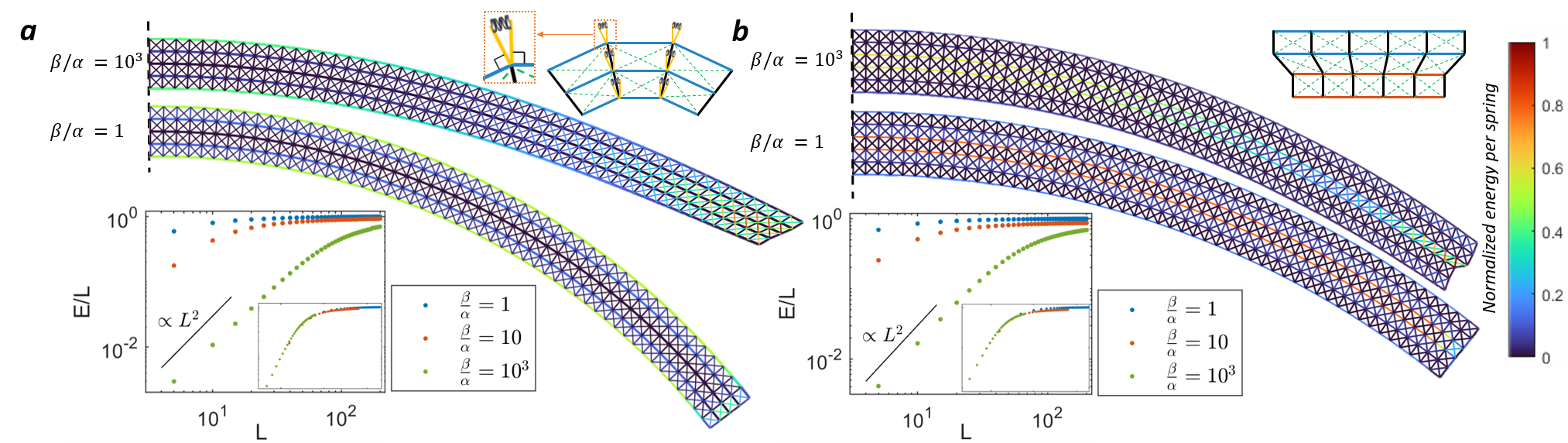}
\caption{Numerical results of spring chains in $2D$ with linear rest-length profile and bend resistance (a), and Timoshenko bi-layer (b). (a) The rest-length of springs per chain (top to bottom) is $1.1$, $1.075$, $1.05$, $1.025$ and $1$. Transverse springs are of rest-length $1$ and the diagonal springs have rest-lengths matching the regular trapezoid formed by the bounding springs at rest. The right half of the minimal conformations for $L=100$  for $\beta/\alpha=10^3$ and $\beta/\alpha=1$ is shown. The color of each spring is indicative of its elastic energy according to the color-bar, normalized to the spring of highest energy in all the simulations. The bottom-left shows the energy per unit length versus the length of the chain for spring constant ratios of $\beta/\alpha=1,10$ and $10^3$, marked as blue, orange and green dots, respectively. The energies are normalized to the maximal energy density. The inset shows the energy densities with the $x$-axis of each curve rescaled by $\sqrt{\beta/\alpha}$. The top-right inset shows a schematic representation of the source of frustration. (b) Results for Timoshenko-like beams with rest-length profile (top to bottom) of $1.05$, $1.05$, $1.05$, $1$, $1$ and $1$, for $L=100$, in the same format as in panel (a).
}
\label{fig:straight_bend}
\end{figure*}

The natural continuum approach for the description of the geometry of such $2D$ elastic systems follows by parameterizing them using longitudinal and transverse coordinates, $u$ and $v$, and examining the resulting metric. For a collection of chains, each with uniform rest-lengths, displaying linear and quadratic variation in the rest-lengths across the chains, the reference metric reads:
\begin{equation*}
  \Bar{a}= \begin{pmatrix}
(1 +k_g v-\frac{k}{2} v^2)^2 & 0\\
0 & 1
\end{pmatrix}.
\end{equation*}
This is the metric of a ribbon whose geometry is characterized by the geodesic curvature and Gaussian curvature of its midline, $v=0$. These assume the values $k_g$ and $k$ and quantify the in-plane turning rate of the ribbon and its non-planarity, respectively. Specifically, any non-vanishing value of the Gaussian curvature, $k$, implies that the reference metric is non-Euclidean and thus has no stress-free embedding in the plane \cite{ESK13}. Such frustration resulting from incompatibility between the reference metric and the embedding space is a common source of frustration \cite{SG05,MPNM14,HAS+19,ZGDS19,THM+22,SHG24}. We capture the mechanical response of the ribbon using a higher-order generalization of the Reddy beam approach \cite{Lev81,HR88}, in which the shear angle is expanded in transverse moments, and traction-free boundary conditions leave as fundamental variables the longitudinally varying arclength, $s(u)$, curvature $\kappa(u)$, the midline shear angle $\theta(u)$, and its first transverse moment, the transverse shear gradient $\theta_v(u)$.  The resulting dimensionally reduced Hamiltonian follows by integrating out the narrow dimension $-t/2\le v \le t/2$. We expand the Hamiltonian to second order in the generalized strains to obtain (for further details see supplementary material \cite{SupplementalMaterial}): 
\begin{equation}\label{ham}
\begin{aligned}
\frac{H}{Et} =& \int_{-\frac{L}{2}}^{\frac{L}{2}}  \,\Bigg[ \,
\frac{2}{15(1+\nu)}\,\theta^{2}+\frac{t^{2}}{24}\left(\kappa+k_{g}+\frac{4}{5}\gamma\right)^{2}+\\
&
 +\frac{t^{2}}{3150}\gamma^{2}+\frac{1}{2}\left(\Delta s-\frac{ t^{2}}{24}\left(\frac{7 }{10}\gamma_{v}-k\right)\right)^{2}+\\
&+\frac{t^{2}}{210(1+\nu)}\,\theta_{v}^{2}
+\frac{t^{4}}{4200}\left(\gamma_{v}-\frac{5}{3}k\right)^{2} 
\Bigg]du,
\end{aligned}
\end{equation}
where $E$ is Young's modulus, $\nu$ is Poisson's ratio, $\Delta s=s-1$, and $\gamma$ and $\gamma_v$ are defined through the linear relations $\gamma = \partial_u\theta$ and $\gamma_v = \partial_u\theta_v$, which can be considered the compatibility conditions, as they tie the gradients of the shear fields to the stretching-related fields they must accommodate. The terms quadratic in $\theta$ and $\theta_v$, carrying the factor $1/(1+\nu)$, originate from the shear strain and are governed by the shear modulus, while the remaining terms originate from the longitudinal strain and are governed by the stretching modulus, in direct analogy with $\alpha$ and $\beta$ of the discrete models. The case of $k=0$ is frustration-free. 
For $k \neq 0$, the frustration is relieved by setting $\gamma_v = \tfrac{5}{3}k$, forcing $\theta_v$ to grow linearly along the beam with a cumulative shear energy scaling as $\frac{t^2}{1+\nu}k^2 L^3$. This solution is overtaken by the uniform one, $\theta_v = 0$, whose stretching energy scales as $t^4 k^2 L$, and the crossing of the two yields $\tilde{L}\simeq\sqrt{\frac{3(1+\nu)}{5}}t $ (see supplementary material \cite{SupplementalMaterial}). The length-scale is thus determined by the ratio of the coefficients of the
last two terms in \eqref{ham}, shear to stretching, in analogy with the previous system.
While for isotropic continuous elasticity this ratio depends only on $\nu$ and $t$, in metamaterials  and small systems in which the lattice spacing is comparable to the narrow dimension, the value that this ratio of longitudinal deformation to shear moduli assumes could be dramatically increased. In the discrete model depicted in Fig. \ref{fig:bend} this ratio is determined by $\beta$ and
$\alpha$ and indeed can be made arbitrarily large. Under this variation, equation \eqref{ham} predicts the correct scaling for $\tilde{L}$ and the energy scaling in the cumulative domain. As the source of frustration in this system corresponds to $k\ne 0$ in the continuum limit, the resulting shear profile is expected to be even about the midline. This is difficult to discern in the discrete model depicted in  Fig. \ref{fig:bend} due to the low amplitude of the frustration. A higher amplitude example is provided in the supplementary material \cite{SupplementalMaterial} and agrees with this prediction. 

For the pure $1D$ conformation considered in the first example, any non-trivial rest-length profile would lead to frustration. In contrast, in the second example linear gradients were accommodated by bending, yet a non-linear rest-length profile would lead to frustration. A distinct class of frustrated systems associated with a linear rest-length profile, compatible with planar geometry, is frustrated as a result of bending resistance of each of the individual filaments.
The bend resistance is localized to the hinges and can arise as a result of structured bonds between the links in the chain, such as in the case of patchy particles \cite{ZG04,SHG22}, or in the hinges in metamaterials \cite{FLT+17,BEIH20}. The linear variation of the rest-lengths favors bending,  while the hinges between the segments favor a straight alignment, leading to a conflict. This frustrated system is characterized by a Euclidean reference metric with $k_g\neq 0$ and $k=0$, and an additional term penalizing bending. The continuum Hamiltonian is similar to equation \eqref{ham} in which $\kappa$ favors a finite value, yet has an additional bending term proportional to $t^2 \kappa^2$ favoring a vanishing $\kappa$. The frustration thus arises from the tension between the tendencies of $\theta$ and $\gamma$, giving rise to a shear profile that is antisymmetric about the midline.
The numerical results of the discrete version of this model are summarized in Fig. \ref{fig:straight_bend}. We denote the shear-related coefficient of the diagonal springs as $\alpha$, and the $\kappa$-related coefficients of the longitudinal and transverse springs and the bend-resisting term as $\beta$. As can be seen, the region in which the cumulative response is observed scales as $\sqrt{\beta/\alpha}$. This scaling remains, although the source of frustration is different from those in the previously discussed models. Notice that the spatial extent over which residual stresses are observed, and the magnitude of these stresses are decoupled. While $\tilde{L}$ here scales similarly to the previous cases, the values assumed by the shear are much larger.

The last frustrated quasi-$1D$ system we consider shares structural motifs with all previous examples, and is arguably the most common. It is composed of two rectangular beams with different longitudinal rest-lengths attached along a common interface, and is known in the continuum limit as the Timoshenko bi-layer. The length mismatch at the interface creates a bending tendency, while the locally flat reference configuration of each beam resists this bend. This class of systems in which stress-free domains are frustrated through their coupling along an interface is common 
in both natural and synthetic systems \cite{Tim25,AEKS11,KO18,Efr20,SDS+20}.   
The numerical results of a discrete version of this system are summarized in panel (b) of Fig. \ref{fig:straight_bend}. As can be seen, even in such a case, where the length mismatch is confined to a single interface, long-lasting longitudinal cumulative frustration can be observed, yet requires soft shear. Macroscopic realizations of the Timoshenko bi-layer do not typically exhibit soft shear and thus will not exhibit cumulative response.
However, in realizations at the microscopic scale, shear deformations may play a key role \cite{SHJ+15,YXC16}. This is expected in particular when the lattice spacing is comparable to the dimensions of the beam and the lattice is cubic, which is prone to zero-modes \cite{HRN18,MHK24}. 
A similar effect was recently observed in metamaterial non-Euclidean trumpets  \cite{WRSG25}. 

Although all of the systems mentioned above demonstrate frustration saturation to a spatially invariant compatible solution, other response modes may also stop the propagation of longitudinal gradients. One such mode is the introduction of defects that locally absorb the frustration and allow the gradients to reset to values closer to their locally favored values \cite{Sel22a,ME22b}. In the context of incommensurate chains of springs, such defects require changing the connectivity of the lattice. In other systems, defects may disrupt the continuity of the gradients without resulting in a local charge, e.g. by only partially connecting the building blocks \cite{THM+22,WG24}. 
These defects are suppressed if the energy associated with their nucleation is higher than that of the saturated compatible structure. 
A different mechanism for reducing the strains associated with gradients is the introduction of a grain boundary along the growth direction, often leading to branching. 
The accumulation of strain at the free edge may cause it to significantly distort to the extent that the resulting interparticle spacing is closer to the native spacing of a different crystalline orientation or to that of another polymorph altogether. As the new grain takes over, longitudinal strains are effectively reset. 

The four systems discussed above are associated with distinct frustration mechanisms that stem from different contradictory tendencies. As a result, they all have different spatially invariant minimal solutions. 
Nevertheless, in all systems $\Tilde{L}\propto \sqrt{\beta/\alpha}$, as their compatibility conditions are of similar order. Note that the increase of $\tilde{L}$ with $\beta/\alpha$ is not specific to this form of compatibility condition. For example, a compatibility condition of second differential order would result in scaling of $\left(\beta/\alpha\right)^\frac{1}{4}$. In addition, if the compatibility condition is nonlinear \cite{NE18,ME22b}, the frustration amplitude may affect $\Tilde{L}$ as well, giving an extra handle for tuning its scale.

In this work, we discuss four distinct systems representing common sources of frustration in quasi-$1D$ systems. In any such system the slenderness of the structure leads to an emergent stiffness in the Hamiltonian along the extended dimension \cite{ME21a}. Since gradients in the transverse directions accumulate along a relatively narrow region, the formation of a gradient along the extended longitudinal dimension becomes prohibitively energetically expensive in comparison. Thus, introducing constitutive softness along the long dimension restores the favorability of gradient formation, recovering the longitudinal cumulative response. This principle could be directly applied to other systems not considered here. For example, the frustration of twisted crystals in $3D$ results from longitudinal length mismatch of material fibers, as the length traveled by the outermost fibers is greater than the inner ones \cite{HBBG16,HAS+19,Gra20a,SDS+20}. In particular it may be considered as a non-Euclidean metric embedded in flat space (resembling the case $k_g=0, k\ne0$ in eq.  \eqref{ham}). This analysis implies that regardless of the details of the underlying geometry, the slenderness suppresses longitudinal cumulative response, yet it may be recovered by soft shear, in agreement with \cite{Gra20a}, opening the door for length-dependent phenomena.

\section*{acknowledgments}
This work was funded by the ISF Grant No. 1444/21 and DFG Grant No. 525814459. We thank Sam Safran for helpful discussions and for bringing Ref.~\cite{FV49} to our attention.

\section*{Data availability}
The code used to generate the numerical results and figures in this article
is openly available on Zenodo \href{https://doi.org/10.5281/zenodo.17948852}{doi:10.5281/zenodo.17948852} .

\bibliography{1D}

\end{document}


\maketitle

\section{Incommensurate chains}

\subsection{Continuum model}

The discrete Hamiltonian reads:
\begin{align*}
H=&\mathop\sum_{i=1}^{N-1}[\frac{\gamma }{2}(X_{i+1}-X_i-l_1)^2+\frac{\beta}{2}(x_{i+1}-x_i-l_0)^2]+\mathop\sum_{i=1}^{N}\frac{\alpha}{2} (X_i-x_i)^2,    
\end{align*}
where $X_i$ ($x_i$) are the locations of the vertices in chains of springs of elastic constants $\gamma$ ($\beta$) and rest-lengths $l_1$ ($l_0$). Each pair ($X_i$,$x_i$) is coupled by a tethering spring of elastic constant $\alpha$ and vanishing rest-length. 
We next transform the variables by defining $\Bar{l}=\frac{\gamma l_1+\beta l_0}{\gamma+\beta}$ and $\eta_i=x_i-X_i$. We divide by $\Bar{l}$:
\begin{align*}
H=&\mathop\sum_{i=1}^{N-1}[\frac{\gamma \Bar{l}^2}{2}\left(\frac{X_{i+1}-X_i}{\Bar{l}}-\frac{l_1}{\Bar{l}}\right)^2+\frac{\beta \Bar{l}^2}{2}\left(\frac{\eta_{i+1}-\eta_i}{\Bar{l}}+\frac{X_{i+1}-X_i}{\Bar{l}}-\frac{l_0}{\Bar{l}}\right)^2]+\mathop\sum_{i=1}^{N}\frac{\alpha}{2}\eta_i^2,    
\end{align*}
By taking the continuum limit of differences divided by $\Bar{l}$ one gets:
\begin{equation*}
H=\int_{-\frac{L}{2}}^{\frac{L}{2}}[\frac{\gamma \Bar{l}}{2}(\frac{dX}{dx}-\frac{l_1}{\Bar{l}})^2+\frac{\beta \Bar{l}}{2}(\frac{d\eta}{dx}+\frac{dX}{dx}-\frac{l_0}{\Bar{l}})^2+\frac{\alpha}{2\Bar{l}}\eta^2]dx.  
\end{equation*}
Minimization using the Euler-Lagrange process yields the following set of equations:
\begin{align*}
  &\frac{\alpha}{\Bar{l}}\eta-\beta \Bar{l} (\eta''+X'')=0, \qquad\gamma \Bar{l} X''+\beta \Bar{l} (\eta''+X'')=0,\\
  &X'(\frac{L}{2})=\frac{l_1}{\Bar{l}}, \quad \eta'(\frac{L}{2})=\frac{l_0-l_1}{\Bar{l}},\quad X(0)=0,\eta(0)=0.
\end{align*}
The solution to these equations is:
\begin{align*}
   \eta(x)&= \frac{ (l_0-l_1) \sinh \left(\frac{\sqrt{\alpha } \sqrt{\frac{1}{\beta} +\frac{1}{\gamma}}}{ \Bar{l}} x\right)}{\sqrt{\alpha }
   \sqrt{\frac{1}{\beta} +\frac{1}{\gamma}} \cosh \left(\frac{\sqrt{\alpha } \sqrt{\frac{1}{\beta} +\frac{1}{\gamma}}}{\Bar{l}}\frac{L}{2}\right)},
   X(x)=x+\frac{ (l_1-l_0)  \sinh \left(\frac{\sqrt{\alpha }  \sqrt{\frac{1}{\beta} +\frac{1}{\gamma}}}{ \Bar{l}} x\right)}{\gamma\sqrt{\alpha }
   \left(\frac{1}{\beta} +\frac{1}{\gamma}\right)^{3/2} \cosh \left(\frac{\sqrt{\alpha }  \sqrt{\frac{1}{\beta} +\frac{1}{\gamma}}}{
   \Bar{l}} \frac{L}{2}\right)}.
\end{align*}

The resulting energy reads:
\begin{equation*}
E=\frac{  (l_0-l_1)^2  L\left(1-\frac{ \tanh \left(\frac{\sqrt{\alpha } L \sqrt{\frac{1}{\beta} +\frac{1}{\gamma}}}{2  \Bar{l}}\right)}{\frac{\sqrt{\alpha } L \sqrt{\frac{1}{\beta} +\frac{1}{\gamma}}}{2  \Bar{l}}}\right)}{2 \Bar{l} (\frac{1}{\beta} +\frac{1}{\gamma} )}.
\end{equation*}

\subsubsection{Compatibility based analysis}
In this analysis, we follow the framework described in \cite{ME21a}. Here we focus on a quasi-$1D$ system where the system can be described by one-dimensional functions with the transverse configuration expressed as an expansion around the midline (or true $1D$ in the current case). A more general use of the analysis, for the case of an isotropic domain in higher dimensions can be found in \cite{ME21a}. The compatibility condition of the two coupled chains in $1D$ in the continuum limit reads $\delta-\xi=\eta'$. This compatibility condition implies that any mismatch between $\delta$ and $\xi$ will be accumulated in $\eta$, as these are not independent. Expanding the variables in spatial orders, i.e. $\delta=\frac{l_0}{\Bar{l}}+\delta_0+\delta_1 x+...$, $\eta=\eta_0+\eta_1 x+...$ and $\xi=\frac{l_1}{\Bar{l}}+\xi_0+\xi_1 x+...$, and substituting the expansions into the compatibility condition yields to leading order:
\begin{equation*}
\frac{l_0}{\Bar{l}}+\delta_0=\eta_1+\frac{l_1}{\Bar{l}}+\xi_0.
\end{equation*}
Thus, for a sufficiently small system, the solution is dominated by a gradient in $\eta$. As long as the system is smaller than the geometric length scale associated with the frustration (to be derived next), the non-uniform solution which is of the lowest energy is obtained by setting $\eta_1=\frac{l_0-l_1}{\Bar{l}}$. Plugging this into the Hamiltonian results in energy scaling of 
\begin{equation*}
    \int_{-L/2}^{L/2}\frac{\alpha}{2\Bar{l}}\frac{(l_0-l_1)^2}{\Bar{l}^2}x^2Dx=\frac{\alpha  (l_0-l_1)^2}{24 \Bar{l}^3}L^3.
\end{equation*}
The uniform-distortion solution, associated with the saturation of the cumulative response, is dominated by the optimal distribution of the frustration between $\delta_0$ and $\xi_0$. In the general case the uniform energy reads
 \begin{equation*}
     \int_{-L/2}^{L/2}\left[\frac{\gamma\bar{l}}{2}\,\xi_0^{\,2}
+\frac{\beta\bar{l}}{2}\left(\xi_0-\frac{l_0-l_1}{\bar{l}}\right)^{\!2}\right]dx
=\frac{\left(l_0-l_1\right)^{2}}{2\,\bar{l}\left(\frac{1}{\beta}+\frac{1}{\gamma}\right)}\,L,
 \end{equation*}
 where the optimal solution is obtained by optimizing with respect to $\xi_0$. Comparing the two energies, the super-extensive scaling is expected to remain valid as long as $L \ll \frac{\Bar{l}}{\sqrt{\alpha(\frac{1}{\beta}+\frac{1}{\gamma})}}$. This distance is the geometric length scale associated with the frustration, related to the validity of the expansion above. These results are in agreement with the explicit solution of the Hamiltonian, shown in the subsection above. 

\subsection{Solution to the discrete model}

We solve the discrete model of two coupled incommensurate chains for the case of $2N+1$ vertices. The general Hamiltonian reads:
\begin{equation*}
H=\mathop\sum_{i=-N}^{N-1}\left[\frac{\gamma }{2}(X_{i+1}-X_i-l_1)^2+\frac{\beta}{2}(x_{i+1}-x_i-l_0)^2\right]+\mathop\sum_{i=-N}^{N}\frac{\alpha}{2} (X_i-x_i)^2.    
\end{equation*}
We start by transforming the variables to the deviations from the na\"ive locations of the $x_i$ particles, meaning $\eta_i=x_i-i \bar{l}$ and $\xi_i=X_i-i \bar{l}$, where $\Bar{l}=\frac{\gamma l_1+\beta l_0}{\gamma+\beta}$. The resulting Hamiltonian reads:
\begin{equation*}
H=\mathop\sum_{i=-N}^{N-1}\left[\frac{\gamma }{2}\left(\xi_{i+1}-\xi_i-\frac{\beta}{\gamma+\beta}\left(l_1-l_0\right)\right)^2+\frac{\beta}{2}\left(\eta_{i+1}-\eta_i +\frac{\gamma}{\gamma+\beta}\left(l_1-l_0
\right)\right)^2\right]+\mathop\sum_{i=-N}^{N}\frac{\alpha}{2} (\xi_i-\eta_i)^2.   
\end{equation*}

We derive a bulk equation for both transformed variables by taking discrete derivatives:
\begin{align*}
    &  \alpha \left(\xi_n-\eta_n\right)+ \beta \left(\eta_{n+1}+\eta_{n-1}-2\eta_{n} \right)=0\\
    &  \alpha \left(\xi_n-\eta_n\right)- \gamma \left(\xi_{n+1}+\xi_{n-1}-2\xi_{n} \right)=0
\end{align*}
Solving this results in recursive relations including 4 constants. \\
To be more explicit, the solution gives 
\[
\gamma \xi(n)+\beta \eta(n) =c_1 +c_2 n+(\gamma-\beta)(c_3 \Lambda^n+c_4 \Lambda^{-n} ), \quad 
\eta(n)-\xi(n)= c_3 \Lambda^n+c_4 \Lambda^{-n} 
\]
where above 
$\Lambda=\left(1+\Delta-\sqrt{\Delta(2+\Delta)}\right)$ and  
$\Delta=\frac{\alpha}{2}(\frac{1}{\beta}+\frac{1}{\gamma})$.
The boundary conditions arise from the terms for $\xi(-N),\xi(N),\eta(-N)$ and $\eta(N)$, where we have $2N+1$ points centered about the origin. $c_1$ is a translational gauge freedom that we can set to zero. The additional boundary conditions give $c_2=0$, and $c_4=-c_3$.  
The full solution is given by 
\[
X_n=n \bar{l}-\frac{\beta}{\gamma+\beta}(l_0-l_1)C\sinh(\lambda n),\qquad
x_n=n \bar{l}+\frac{\gamma}{\gamma+\beta}(l_0-l_1)C \sinh(\lambda n),
\]
where $\lambda=\log(\Lambda)$ and $C=1/(\sinh(\lambda (N+1))-\sinh(\lambda N))$.

\subsection{Simulations of discrete model}

The general Hamiltonian reads:
\begin{equation*}
H=\mathop\sum_{i=1}^{M}\mathop\sum_{j=1}^{N-1}\frac{\beta}{2}(x_{i,j+1}-x_{i,j}-l_i)^2+\mathop\sum_{i=1}^{M-1}\mathop\sum_{j=1}^{N}\frac{\alpha}{2} (x_{i,j}-x_{i+1,j})^2.    
\end{equation*}
where $M$ is the number of chains, $N$ is the number of vertices per chain, $x_{i,j}$ is the j'th vertex in the i'th chain, $l_i$ is the spring rest-length at the i'th chain, and $\alpha$ and $\beta$  are spring constants of inter-chain and intra-chain springs, respectively.  

The simulations were performed using \textit{MATLAB} \cite{MATLAB}. The conformations of minimal energy were found using \textit{fminunc} with step tolerance of $10^{-9}$ and optimality tolerance of $10^{-8}$. The initial conformation was set to equal spacing in all chains of the mean value of the rest-lengths in all the chains, $l_{mean}$. The spring constant within the chains was set to $\beta=\frac{10^4}{\frac{1}{2}\mathop\sum_{i=1}^{M}\left( l_i-l_{mean}\right)^2}$ and the tethering inter-chain springs were set to $\alpha=\beta$,  $ 10^{-1}\beta$ and  $10^{-3}\beta$.  The simulations studying the width of the boundary layers versus the number of chains were performed in a similar manner, with spring constants of $\beta=\alpha=10^3$. The longitudinal rest length was symmetric around the central chain with the rest length difference between adjacent chains being $\Delta l=0.0025$. The shortest  rest length of the springs, of the central chain, was set to $1$.

\section{Generalized shearable beam}

We derive a  $2D$ shearable beam model, using a higher order generalization of the Reddy-beam approach,  in which the shear angle is expanded in transverse moments, and the transverse normal stretch is described by a field. We derive the quadratic Hamiltonian \(H^{(2)}\).
Let \(u\in[-L/2,L/2]\) denote the longitudinal coordinate and \(v\in[-t/2,t/2]\) the transverse coordinate. The midline of the beam is described by \(\mathbf r(u)\), with orthonormal tangent and normal unit-vectors \(\hat{\mathbf t}(u)\) and \(\hat{\mathbf n}(u)\). We write
\begin{equation}
\partial_u \mathbf r=(1+\Delta s)\,\hat{\mathbf t},
\label{eq:r_u}
\end{equation}
where \(\Delta s(u)\) is the longitudinal stretch of the midline.
The tangent and normal satisfy
\begin{equation}
\partial_u \hat{\mathbf t}=(1+\Delta s)\kappa\,\hat{\mathbf n},
\qquad
\partial_u \hat{\mathbf n}=-(1+\Delta s)\kappa\,\hat{\mathbf t},
\label{eq:tn_u}
\end{equation}
where \(\kappa(u)\) is the curvature of the midline.
Since we compute only the first-order strains, required for the second-order Hamiltonian, products of the fields in the expansion will be dropped systematically.

\subsection{Kinematic ansatz}

We introduce a transverse stretch field \(\psi(u,v)\) and a shear-angle profile \(q(u,v)\), and define the deformation through
\begin{equation}
\partial_v \mathbf R(u,v)
=
\bigl(1+\psi(u,v)\bigr)\,\eta\!\bigl(q(u,v)\bigr)\,\hat{\mathbf n}(u),
\label{eq:R_v_def_first}
\end{equation}
where \(\eta(\alpha)\) is the planar rotation by angle \(\alpha\).
We expand the shear angle in transverse moments as
\begin{equation}
q(u,v)
=
\theta_0(u)
+v\,\theta_1(u)
+\frac{v^2}{2}\theta_2(u)
+\frac{v^3}{6}\theta_3(u),
\label{eq:q_moments}
\end{equation}
and the transverse stretch as
\begin{equation}
\psi(u,v)
=
\psi_0(u)
+v\,\psi_1(u)
+\frac{v^2}{2}\psi_2(u)
+\frac{v^3}{6}\psi_3(u)
+\frac{v^4}{24}\psi_4(u).
\label{eq:psi_moments}
\end{equation}
The higher moments in the shear angle are introduced as they allow to lower the elastic energy associated with the linear and quadratic profiles of the reference metric. The expansions of $q$ and $\psi$ are truncated at these orders because they are the minimal generalizations that can comply with the boundary conditions that are applied later in this derivation.
To obtain the first-order strains, it is sufficient to linearize \eqref{eq:R_v_def_first}. Since
$\eta(q)\hat{\mathbf n}=-\sin q\,\hat{\mathbf t}+\cos q\,\hat{\mathbf n}$,
$\sin q=q+O(q^3)$ and $\cos q=1+O(q^2)$, we have, to first order,
\begin{equation}
\partial_v \mathbf R
=
-q\,\hat{\mathbf t}
+
(1+\psi)\,\hat{\mathbf n}.
\label{eq:R_v_linear}
\end{equation}
Imposing \(\mathbf R(u,0)=\mathbf r(u)\), the integration of \eqref{eq:R_v_linear} gives
\begin{equation*}
\mathbf R(u,v)
=
\mathbf r(u)-Q(u,v)\,\hat{\mathbf t}(u)+\bigl(v+P(u,v)\bigr)\,\hat{\mathbf n}(u),
\label{eq:R_decomp_linear}
\end{equation*}
where
\begin{equation*}
Q(u,v):=\int_0^v q(u,\xi)\,d\xi,
\qquad
P(u,v):=\int_0^v \psi(u,\xi)\,d\xi.
\label{eq:Q_P_defs}
\end{equation*}
Using \eqref{eq:q_moments} and \eqref{eq:psi_moments},
\begin{equation*}
Q(u,v)
=
v\theta_0
+\frac{v^2}{2}\theta_1
+\frac{v^3}{6}\theta_2
+\frac{v^4}{24}\theta_3,
\label{eq:Q_explicit_linear}
\end{equation*}
and
\begin{equation*}
P(u,v)
=
v\psi_0
+\frac{v^2}{2}\psi_1
+\frac{v^3}{6}\psi_2+\frac{v^4}{24}\psi_3(u)+\frac{v^5}{120}\psi_4(u).
\label{eq:P_explicit_linear}
\end{equation*}
Differentiating with respect to \(u\),
\[
\partial_u \mathbf R
=
\partial_u\mathbf r
-Q_u\,\hat{\mathbf t}
-Q\,\partial_u \hat{\mathbf t}
+P_u\,\hat{\mathbf n}
+(v+P)\,\partial_u\hat{\mathbf n}.
\]
Using \eqref{eq:r_u} and \eqref{eq:tn_u}, and dropping products of the fields (leading to second order terms), we obtain
\begin{equation}
\partial_u \mathbf R
=
\bigl(1+\Delta s-Q_u-\kappa v\bigr)\,\hat{\mathbf t}
+
P_u\,\hat{\mathbf n}.
\label{eq:R_u_linear}
\end{equation}
The derivatives of \(Q\) and \(P\) are
\begin{equation}
Q_u(u,v)
=
v\theta_0'
+\frac{v^2}{2}\theta_1'
+\frac{v^3}{6}\theta_2'
+\frac{v^4}{24}\theta_3',
\label{eq:Q_u_linear}
\end{equation}
and
\begin{equation}
P_u(u,v)
=
v\psi_0'+\frac{v^2}{2}\psi_1'+\frac{v^3}{6}\psi_2'+\frac{v^4}{24}\psi_3'+\frac{v^5}{120}\psi_4'.
\label{eq:P_u_linear}
\end{equation}

\subsection{Metric expanded to first order}

The metric is
\[
a_{\alpha\beta}=\partial_\alpha \mathbf R\cdot \partial_\beta \mathbf R.
\]
Since we need only the first-order strains, we compute only the first-order metric.

\paragraph{The component \(a_{uu}\).}
From \eqref{eq:R_u_linear},
\[
a_{uu}
=
\bigl(1+\Delta s-Q_u-\kappa v\bigr)^2+(P_u)^2.
\]
Since \((P_u)^2\) is second order, it does not contribute to \(a_{uu}^{(1)}\). 
Using \eqref{eq:Q_u_linear},
\begin{equation}
a_{uu}^{(1)}
=
1+2\Delta s
-2v(\kappa+\theta_0')
-v^2\theta_1'
-\frac{v^3}{3}\theta_2'
-\frac{v^4}{12}\theta_3'.
\label{eq:auu_first}
\end{equation}

\paragraph{The component \(a_{uv}\).}
Using \eqref{eq:R_u_linear} and \eqref{eq:R_v_linear},
\[
a_{uv}
=
\partial_u\mathbf R\cdot\partial_v\mathbf R
=
-\bigl(1+\Delta s-Q_u-\kappa v\bigr)q+P_u(1+\psi).
\]
Keeping only first-order terms, and substituting \eqref{eq:q_moments} and \eqref{eq:P_u_linear} gives
\begin{equation}
a_{uv}^{(1)}
=
-\left(
\theta_0
+v\theta_1
+\frac{v^2}{2}\theta_2
+\frac{v^3}{6}\theta_3
\right)
+
\left(
v\psi_0'
+\frac{v^2}{2}\psi_1'
+\frac{v^3}{6}\psi_2'
+\frac{v^4}{24}\psi_3'
+\frac{v^5}{120}\psi_4'
\right).
\label{eq:auv_first}
\end{equation}

\paragraph{The component \(a_{vv}\).}
From \eqref{eq:R_v_linear},
\[
a_{vv}
=
q^2+(1+\psi)^2.
\]
Since \(q^2\) is second order, it does not contribute to \(a_{vv}^{(1)}\).
Using \eqref{eq:psi_moments},
\begin{equation}
a_{vv}^{(1)}
=
1+2\psi_0+2v\psi_1+v^2\psi_2+\frac{v^3}{3}\psi_3(u)+\frac{v^4}{12}\psi_4(u).
\label{eq:avv_first}
\end{equation}

\subsection{Reference metric and first-order strains}

We take the non-Euclidean reference metric in the form
\begin{equation}
  \bar{a}= \begin{pmatrix}
(1 +k_g v-\frac{k}{2} v^2)^2 & 0\\
0 & 1
\end{pmatrix}
\label{eq:reference_metric}.
\end{equation}
where \(k_g\) and \(k\) are the prescribed geometric parameters, determining the reference geodesic and Gaussian curvatures at the midline, respectively. These geometric parameters are considered to be small (on the same order of the fields), as discussed later.
The strain tensor is defined as
\begin{equation}
e_{\alpha\beta}
=
\frac12(a_{\alpha\beta}-\bar a_{\alpha\beta}).
\end{equation}
At the order relevant for the quadratic Hamiltonian, only the first-order strains are needed:

\paragraph{Longitudinal strain.}
Using \eqref{eq:auu_first} and \eqref{eq:reference_metric},
\begin{equation}
e_{uu}^{(1)}
=
\Delta s
-v(\kappa+\theta_0'+k_g)
-\frac{v^2}{2}(\theta_1'-k)
-\frac{v^3}{6}\theta_2'
-\frac{v^4}{24}\theta_3'.
\label{eq:euu_first_final}
\end{equation}

\paragraph{Mixed strain.}
Using \eqref{eq:auv_first},
\begin{equation}
e_{uv}^{(1)}
=
-\frac12\left(
\theta_0
+v\theta_1
+\frac{v^2}{2}\theta_2
+\frac{v^3}{6}\theta_3
\right)
+\frac12\left(
v\psi_0'
+\frac{v^2}{2}\psi_1'
+\frac{v^3}{6}\psi_2'
+\frac{v^4}{24}\psi_3'
+\frac{v^5}{120}\psi_4'
\right).
\label{eq:euv_first_final}
\end{equation}

\paragraph{Transverse  strain.}
Using \eqref{eq:avv_first},
\begin{equation}
e_{vv}^{(1)}
=
\psi_0+v\psi_1+\frac{v^2}{2}\psi_2+\frac{v^3}{6}\psi_3+\frac{v^4}{24}\psi_4.
\label{eq:evv_first_final}
\end{equation}

\subsection{Quadratic Hamiltonian}

Since the elastic energy is quadratic in the strain, the second-order Hamiltonian \(H^{(2)}\) depends only on the first-order strains \eqref{eq:euu_first_final}--\eqref{eq:evv_first_final}. For a planar $2D$ isotropic beam in 'flatland' \cite{SN88}, the quadratic Hamiltonian is
\begin{equation*}
H^{(2)}
=
\int_{-\frac{L}{2}}^\frac{L}{2}\int_{-t/2}^{t/2}
\left[
\frac{E}{2(1-\nu^2)}
\Bigl((e_{uu}^{(1)})^2+(e_{vv}^{(1)})^2+2\nu e_{uu}^{(1)}e_{vv}^{(1)}\Bigr)
+
\frac{E}{1+\nu}(e_{uv}^{(1)})^2
\right]\,dv\,du.
\label{eq:H2_general}
\end{equation*}
where $E$ is Young's modulus and $\nu$ is Poisson's ratio in $2D$.

\subsection{First-order free-boundary conditions and reduced quadratic Hamiltonian}

We next apply the free-boundary conditions at the same order as the strains, namely at first order in the kinematics, consistent with the quadratic Hamiltonian \(H^{(2)}\).

\subsubsection{Constitutive law and traction-free boundaries}

The first-order stresses for $2D$ isotropic materials are
\begin{equation*}
\sigma_{uu}^{(1)}=\frac{E}{1-\nu^2}(e_{uu}^{(1)}+\nu e_{vv}^{(1)}),
\qquad
\sigma_{vv}^{(1)}=\frac{E}{1-\nu^2}(e_{vv}^{(1)}+\nu e_{uu}^{(1)}),
\qquad
\sigma_{uv}^{(1)}=\frac{2E}{1+\nu}e_{uv}^{(1)}.
\label{eq:first_order_stresses}
\end{equation*}

The side boundaries \(v=\pm t/2\) are traction free, so
\begin{equation*}
\sigma_{vv}^{(1)}(u,\pm t/2)=0,
\qquad
\sigma_{uv}^{(1)}(u,\pm t/2)=0.
\label{eq:traction_free_first_order}
\end{equation*}
Equivalently,
\begin{equation*}
e_{vv}^{(1)}(u,\pm t/2)+\nu e_{uu}^{(1)}(u,\pm t/2)=0,
\qquad
e_{uv}^{(1)}(u,\pm t/2)=0.
\label{eq:first_order_BCs}
\end{equation*}

\subsubsection{Tangential traction condition $e_{uv}^{(1)}(u,\pm t/2)=0$}

From \eqref{eq:euv_first_final},
\begin{equation*}
e_{uv}^{(1)}
=
-\frac12\left(
\theta_0
+v\theta_1
+\frac{v^2}{2}\theta_2
+\frac{v^3}{6}\theta_3
\right)
+\frac12\left(
v\psi_0'
+\frac{v^2}{2}\psi_1'
+\frac{v^3}{6}\psi_2'
+\frac{v^4}{24}\psi_3'
+\frac{v^5}{120}\psi_4'
\right).
\label{eq:euv_first_recalled}
\end{equation*}
Evaluating at \(v=\pm t/2\), the condition \(e_{uv}^{(1)}(u,\pm t/2)=0\) yields
\begin{equation*}
-\left(
\theta_0+\frac{t}{2}\theta_1+\frac{t^2}{8}\theta_2+\frac{t^3}{48}\theta_3
\right)
+
\left(
\frac{t}{2}\psi_0'
+\frac{t^2}{8}\psi_1'
+\frac{t^3}{48}\psi_2'
+\frac{t^4}{384}\psi_3'
+\frac{t^5}{3840}\psi_4'
\right)
=0,
\label{eq:euv_plus}
\end{equation*}
\begin{equation*}
-\left(
\theta_0-\frac{t}{2}\theta_1+\frac{t^2}{8}\theta_2-\frac{t^3}{48}\theta_3
\right)
+
\left(
-\frac{t}{2}\psi_0'
+\frac{t^2}{8}\psi_1'
-\frac{t^3}{48}\psi_2'
+\frac{t^4}{384}\psi_3'-\frac{t^5}{3840}\psi_4'
\right)
=0.
\label{eq:euv_minus}
\end{equation*}
Adding and subtracting these equations gives
\begin{equation*}
\theta_0+\frac{t^2}{8}\theta_2=\frac{t^2}{8}\psi_1'+\frac{t^4}{384}\psi_3',
\label{eq:theta0_theta2_relation2}
\end{equation*}
\begin{equation*}
\theta_1+\frac{t^2}{24}\theta_3=\psi_0'+\frac{t^2}{24}\psi_2'+\frac{t^5}{1920}\psi_4'.
\label{eq:theta1_theta3_relation2}
\end{equation*}

\subsubsection{Normal traction condition $\sigma_{vv}^{(1)}=0$}

 We now apply the normal traction condition, $\sigma_{vv}^{(1)}=\frac{E}{1-\nu^2}\left(e_{vv}^{(1)}+\nu e_{uu}^{(1)}\right)=0$.  Due to the slenderness of the beam we assume, similarly to the Kirchhoff-Love assumption, that the normal condition is enforced point-wise throughout the thickness of the beam, and not just at the boundaries. The terms in the condition must be balanced for any $v$, yielding:
 \begin{equation*}
     \psi_0=-\nu \Delta s,\quad  \psi_1=\nu (\kappa+\theta_0'+k_g),\quad \psi_2=\nu (\theta_1'-k), \quad \psi_3=\nu \theta_2' \quad and \quad \psi_4=\nu \theta_3'.
 \end{equation*}

\subsubsection{Asymptotic ordering}
The expansion so far assumed na\"ively that all the considered fields are small, and should be kept to first order at most. We next assume that the fields vary along the longitudinal direction with some length scale $\bar{l}\gg t$, otherwise the solution must decay to a uniform non-cumulative solution near the ends of the beam. Thus, any longitudinal derivative scales as $\partial_u\sim O(\frac{1}{\bar{l}})$. From the normal boundary condition we learn that the $\psi$ fields scale (at most), as the order of $\Delta s$, $\kappa$, the constants from $\bar{a}$ and the derivatives of the shear-related fields. Thus, the order of the longitudinal derivatives of $\psi$ is at least order $ O(\frac{1}{\bar{l}})$ smaller than the order of the shear-related fields. The tangential boundary conditions are reduced to:
\begin{equation}
\theta_0+\frac{t^2}{8}\theta_2=0 \rightarrow \theta_2=-\frac{8}{t^2}\theta_0,
\label{eq:theta0_theta2_relation}
\end{equation}
\begin{equation}
\theta_1+\frac{t^2}{24}\theta_3=0 \rightarrow \theta_3=-\frac{24}{t^2}\theta_1 .
\label{eq:theta1_theta3_relation}
\end{equation}
Since this condition holds for any $u$ and from continuity of the fields, these relations also hold for their $u$ derivatives. The strains then read:
\begin{equation}
e_{uu}^{(1)}
=
\Delta s
-v(\kappa+(1-\frac{4 v^2}{3 t^2})\theta_0'+k_g)
-\frac{v^2}{2}((1-\frac{2v^2}{t^2})\theta_1'-k),
\label{eq:euu_first_final2}
\end{equation}
\begin{equation}
e_{uv}^{(1)}
=
-\frac{1}{2}\left(
(1-\frac{4v^2}{t^2})\theta_0
+v(1-\frac{4v^2}{t^2})\theta_1
\right),
\label{eq:euv_first_final2}
\end{equation}
\begin{equation}
e_{vv}^{(1)}
=-\nu e_{uu}^{(1)}.
\label{eq:evv_first_final2}
\end{equation}

\subsubsection{Quadratic Hamiltonian}
Substituting \eqref{eq:euu_first_final2}-\eqref{eq:evv_first_final2}, integrating along the narrow dimension, $v$, the quadratic Hamiltonian, expanded to second order up to additive constants reads:

\begin{equation}
\begin{aligned}
H^{(2)} =&Et \int_{-L/2}^{L/2}  \,\Bigg[ \,
\frac{2}{15(1+\nu)}\,\theta_{0}(u)^{2}+\frac{t^{2}}{24}\left(\kappa(u)+k_{g}+\frac{4}{5}\theta_{0}'(u)\right)^{2}
+\frac{t^{2}}{3150}\bigl(\theta_{0}'(u)\bigr)^{2} \\
&+\frac{t^{2}}{210(1+\nu)}\,\theta_{1}(u)^{2}+\frac{1}{2}\left(\Delta s(u)-\frac{ t^{2}}{24}\left(\frac{7 }{10}\theta_{1}'(u)-k\right)\right)^{2}
+\frac{t^{4}}{4200}\left(\theta_{1}'(u)-\frac{5}{3}k\right)^{2} 
\Bigg]du.
\end{aligned}
\label{eq:ham_final}
\end{equation}
In the main text we use $\theta_0=\theta$, $\theta_0'=\gamma$, $\theta_1=\theta_v$, and $\theta_1'=\gamma_v$.  
\subsubsection{Compatibility based analysis}
The frustration is introduced into this Hamiltonian through $k$, favoring non-vanishing $\gamma_v$. Thus, the two relevant terms in \eqref{eq:ham_final} that do not vanish identically at the ground state read:
\begin{equation*}
\begin{aligned}
H^{(2)} =&Et \int_{-L/2}^{L/2}  \,\Bigg[ \,
\frac{t^{2}}{210(1+\nu)}\,\theta_{v}(u)^{2}
+\frac{t^{4}}{4200}\left(\gamma_v(u)-\frac{5}{3}k\right)^{2} 
\Bigg]du.
\end{aligned}
\end{equation*}
Setting $\gamma_v=\frac{5}{3}k$, yields cumulative energy scaling of:
\begin{equation*}
    \int_{-L/2}^{L/2}  
\frac{Et^{3}}{210(1+\nu)}\left(\frac{5}{3}ku\right)^{2}
du = \frac{5Et^{3}}{4536(1+\nu)}k^2L^3.
\end{equation*}
The energy of the non-cumulative uniform solution reads:
\begin{equation*}
    \int_{-L/2}^{L/2}
\frac{Et^{5}}{4200}\left(\frac{5}{3}k\right)^{2} 
du=\frac{Et^5}{1512}k^2L
\end{equation*}
Their crossing defines the length scale associated with frustration at $\Tilde{L}\simeq\sqrt{\frac{3(1+\nu)}{5}}t$.

\section{ 2-dimensional simulations - general method}

This section refers to all the $2D$ simulations. Specific details for each simulation are listed in the separate subsections below. The simulations were performed using \textit{MATLAB} \cite{MATLAB}. The conformations of minimal energy were found using \textit{fminunc} with step tolerance of $10^{-12}$, function tolerance of $10^{-12}$ and optimality tolerance of $10^{-10}$.
The Hamiltonian of the discrete model reads:

\begin{align*}
H=&\mathop\sum_{i=1}^{M}\mathop\sum_{j=1}^{N}\frac{\beta}{2}[|\Vec{r}_{i,j}-\Vec{r}_{i,j+1}|-l_{i}]^2+\mathop\sum_{i=1}^{M-1}\mathop\sum_{j=1}^{N+1}\frac{\beta}{2}(|\Vec{r}_{i,j}-\Vec{r}_{i+1,j}|-l_{t})^2\\  
&+\mathop\sum_{i=1}^{M-1}\mathop\sum_{j=1}^{N}\frac{\alpha}{2}[(|\Vec{r}_{i,j}-\Vec{r}_{i+1,j+1}|-l_{d,i})^2+(|\Vec{r}_{i+1,j}-\Vec{r}_{i,j+1}|-l_{d,i})^2].    
\end{align*}
$||$ symbol represents the norm of the vector. There are $M$ chains of $N$ springs,  $\Vec{r}_{i,j}$ is the location of the vertex $j$ in the chain $i$, $l_i$ is the base rest-length of the longitudinal springs, $l_t$ is the rest-length of the transverse springs, $\beta$ is the elastic constant of both the longitudinal and transverse springs, $\alpha$ is the elastic constant of the diagonal springs and $l_{d,i}$ is the rest-length of the diagonal springs between the $i$ and $i+1$ layers. The last term is added to control the shear modulus of the system and the rest-length is chosen to match the diagonal of a regular trapezoid formed by the four bounding springs at their rest configuration.
In addition, there is an extra term (not written explicitly for brevity) that penalizes inversions of triangles composed of a lateral spring, transverse spring and a diagonal spring with respect to their reference ordering. In case of an inversion (which was not observed in the minimal configurations), a value of $10^{12}$ was added to the energy of the configuration.

\subsection{Non-Euclidean beam}\label{ne}

The simulations were run using $M=5$, $N$ varying between $5$ and $500$ with steps of $5$, longitudinal rest-lengths (top to bottom) of $1.01$, $1.0056$, $1.0025$, $1.0006$ and $1$, lateral rest-length $l_t=1$, $\beta=10^6$ and $\alpha=10^6$, $10^5$ and $10^3$.  

The initial conformations for the case of equal spring constants were computed by finding the mean curvature of the midline and spacing along the chains and in the transverse direction for a minimal conformation of $30$ vertices per chain, initialized as a straight beam with mean rest-length spacing along the longitudinal direction and with rest-transverse-spacing. The resulting minimal conformation was used as the initial configurations for the two other sets of spring constants. In all runs a noise of Gaussian distribution with $0$ mean and standard deviation of $0.001$ was added to the initial condition.  

\begin{figure*}[htbp]
\includegraphics[width=16.5cm]{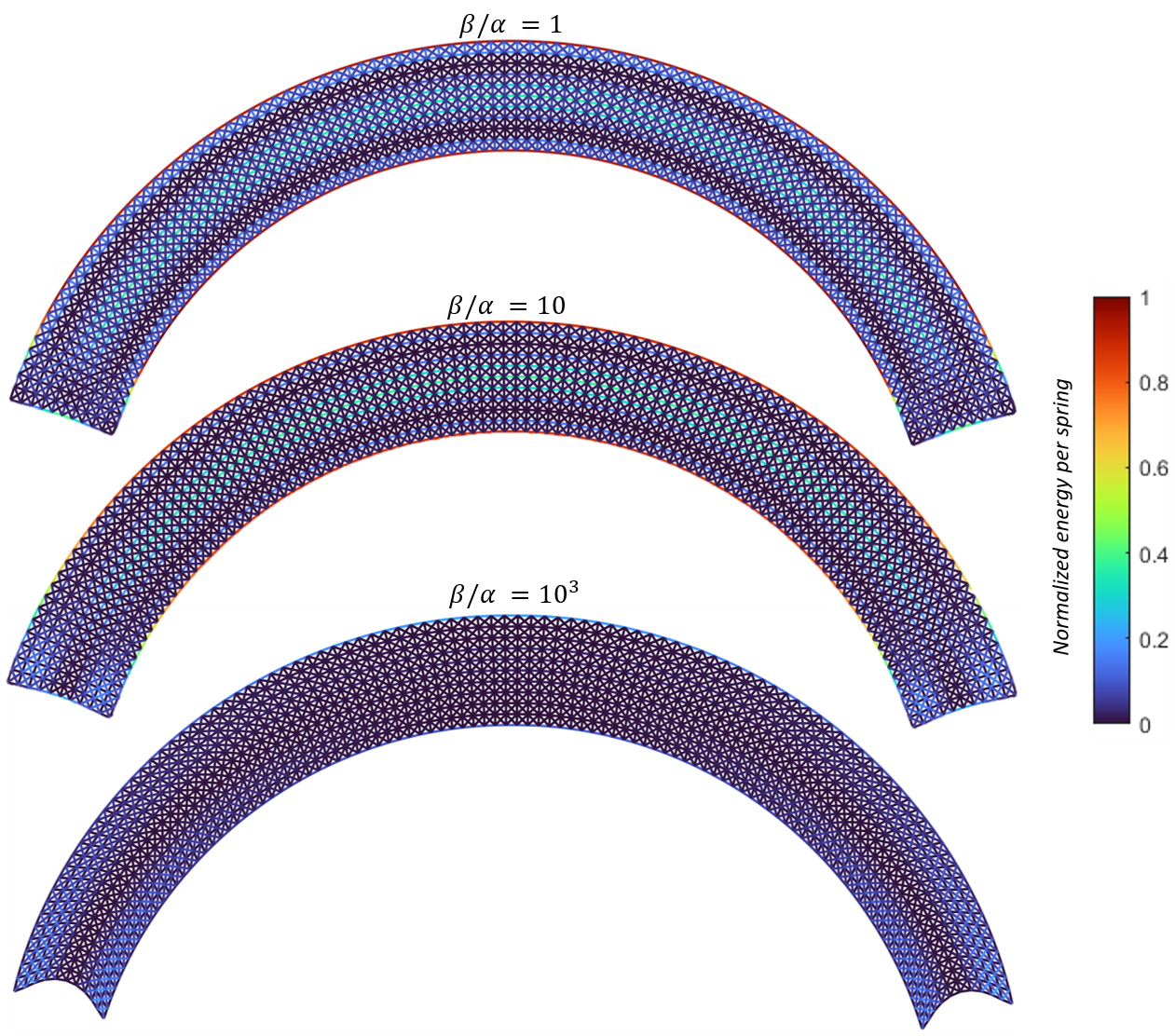}
\caption{Non-Euclidean spring chains restricted to the plane. An example of a wider system, with $11$ chains, and a higher frustration amplitude, highlighting the shear profile in the cumulative regime. This figure is complementary to Fig. $2$ in the main text. The numerical results are obtained for a rest-length profile per chain (bottom to top) of $\bar{l}_{i}=1+0.25\frac{(i-1)^2}{10^2}$. Transverse springs are of rest-length $1$ and the diagonal springs have rest-lengths matching the regular trapezoid formed by the bounding springs at rest. The conformations are for $L=100$ at $\beta/\alpha=1,10$ and $10^3$. The color of each spring is indicative of its elastic energy according to the color-bar, normalized to the spring of highest energy of the three simulations. }
\label{fig:bend}
\end{figure*}

\subsection{Bend-resisting beam}

In the case of the bend-resisting beam, an extra term was added to the Hamiltonian of the elastic springs shown above, accounting for bend resistance. Its form reads:
\begin{equation*}
    k_{bend}\mathop\sum_{i=1}^{M}\mathop\sum_{j=1}^{N-1}(1-\hat{t}_{i,j} \cdot \hat{t}_{i,j+1}),
\end{equation*}
where $k_{bend}$ is the modulus resisting the bend and $\hat{t}_{i,j}$ is the normalized vector that connects the vertices $j$ and $j+1$ in the chain $i$. The simulations were carried out using $M=5$, $N$ varying between $5$ and $200$ with steps of $5$, longitudinal rest-lengths (top to bottom) of $1.1$, $1.075$, $1.05$, $1.025$ and $1$, lateral rest-length $l_t=1$, $\beta=10^4$, $\alpha=10^4$, $10^3$ and $10^1$ and $k_{bend}=\beta=10^4$.

The initial conformations were computed by finding the mean curvature of the midline and spacing along the chains and in the transverse direction for a minimal conformation of $30$ vertices per chain for the case of equal moduli, initialized as a straight beam with mean rest-length spacing along the longitudinal direction and with rest-transverse-spacing. In all runs, Gaussian noise with $0$ mean and standard deviation of $0.025$ was added to the initial condition. In the case of soft-shear ($\alpha=10$), the result of the optimization process was fed as an initial condition to a second round of optimization. 

\subsection{Timoshenko bi-layer}

The simulations were carried out using $M=6$, $N$ varying between $5$ and $200$ with steps of $5$, longitudinal rest-lengths (top to bottom) of $1.05$, $1.05$, $1.05$, $1$, $1$ and $1$, lateral rest-length $l_t=1$, $\beta=10^6$ and $\alpha=10^6$, $10^5$ and $10^3$.  

The initial conformations for the case of equal spring constants were computed by finding the mean curvature of the midline and spacing along the chains and in the transverse direction for a minimal conformation of $30$ vertices per chain, initialized as a straight beam with mean rest-length spacing along the longitudinal direction and with rest-transverse-spacing. The resulting minimal conformation was used as the initial configuration for the two other sets of spring constants. In all runs, a Gaussian noise with $0$ mean and standard deviation of $0.001$ was added to the initial condition.  

\bibliographystyle{unsrt}
\bibliography{1d}